# Ergo - a programming language for Smart Legal Contracts

**Niall Roche**[†**‡*] **Walter Hernandez**[‡*] **Eason Chen**[*] **Jérôme Siméon**[***] **Dan Selman**[***]
[*]Accord Project [**]Docusign [†]University College London [‡]Mishcon de Reya
niall.roche@mishcon.com

## Abstract

We present a smart legal contract platform to support a wide range of smart legal contract use cases. We see this as a step towards improving existing approaches to representing the complexity of legal agreements and executing aspects of these agreements.

The smart contract is a coded computer program that will automatically execute when something triggers it. In contrast, the smart legal contract is a legal agreement in digital and executable code that connects terms and can interact with other software systems.

Clack et al. (2016) provides an encompassing definition of *smart contract* and *smart legal contract* by considering the operational aspect and legal focus of both while basing the definition in the topics of automation and enforceability:

"A smart contract is an automatable and enforceable agreement. Automatable by computer, although some parts may require human input and control. Enforceable either by legal enforcement of rights and obligations or via tamper-proof execution of computer code."

## 1 Introduction

We define a Smart Legal Contract as a human-readable and machine-readable agreement that is digital, consisting of natural language and computable components.

The human-readable nature of the document ensures that signatories, lawyers, contracting parties and others are able to understand the contract.

The machine-readable nature of the document enables it to be interpreted and executed by computers, making the document "smart".

A Smart Legal Contract abstracts the Legal Agreement and executable clauses contained within the agreement.

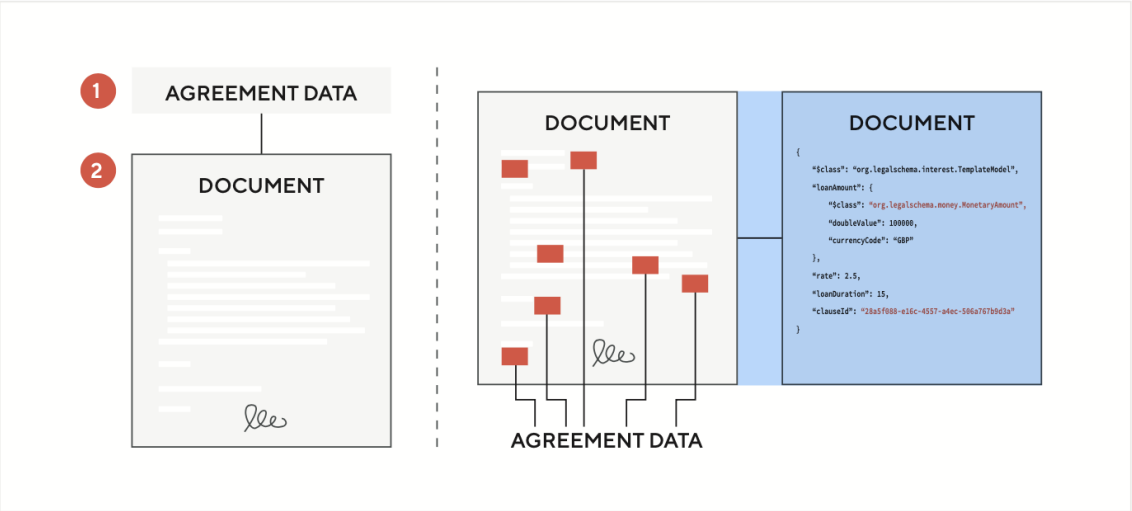

Figure 1: Smart Legal Contract Document Structure [1]

Processes inputs and outputs obligations such as payments due.

A Smart Legal Contract can receive inputs from external data sources via trusted 'Oracles' including:

1. IoT Sensors
2. Payment Providers
3. External Data feeds
4. Application Programming Interfaces (APIs)

The Accord Project consists of three main modules: Cicero, Concerto, and Ergo. The relationship between these modules is outlined in Figure 2 with examples of each concept outlined in Figure 3.

**Cicero** provides a general format to structure rich-text formatted templates for clauses, sections, or contracts as a machine-readable format that can be searched, analyzed, and executed. A number of examples templates have been contributed to a template library. [3]

**Concerto** is a domain specific object modelling language for defining and managing models of both human-readable and machine-readable smart contracts. Concerto has the ability to represent complex data types and a type hierarchy and there are a number of preexisting data models[4] available.

[3]https://templates.accordproject.org
[4]https://models.accordproject.org

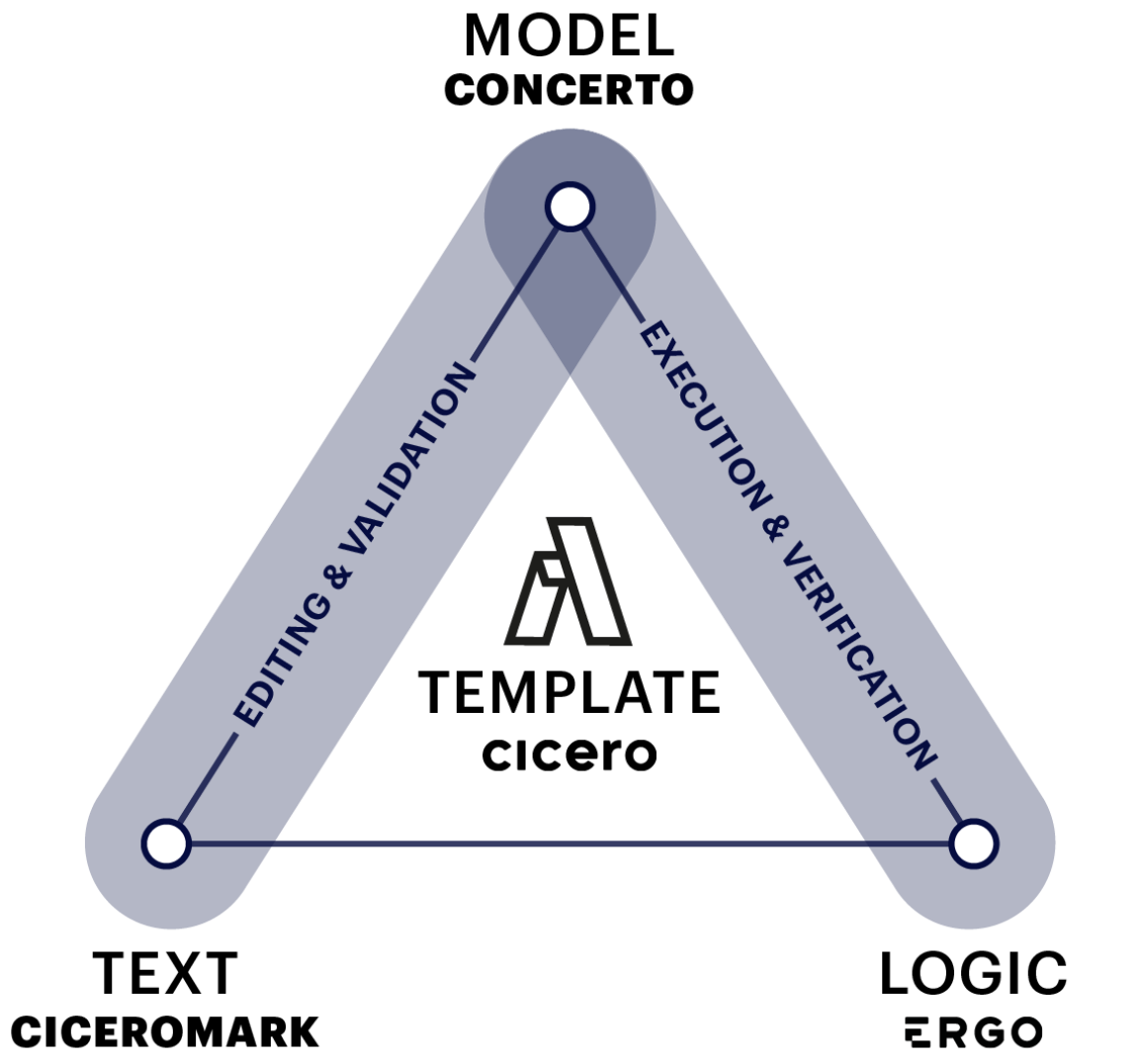

Figure 2: The relationship between Cicero, Concerto, and Ergo [2]

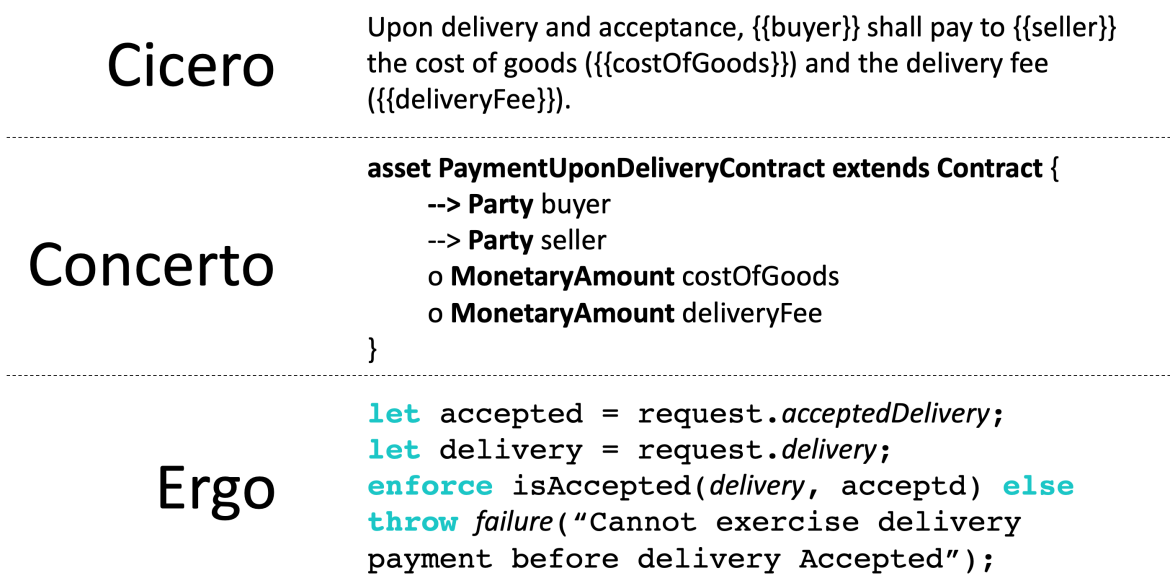

Figure 3: Example of Accord's Template Contract [5]

**Ergo** is a domain-specific language (DSL) that captures the execution logic of legal contracts. Ergo is intended to be accessible to both Lawyers and Developers, helping ensure that the executable logic and the corresponding legal prose are always semantically consistent.

These three modules are intertwined: Cicero utilizes Concerto to express variables in natural language that may be bound to Ergo for logic execution. as outlined in Figure. 2.

Due to the modular architecture of the Ergo compiler Ergo smart legal contracts created using the Accord Project can compiled to JavaScript, Java or WebAssembly, allowing deployment in a Node.js runtime[6] a serverless deployment using the centralised ledger service QLDB[7] from Amazon AWS [8] and on a range of permissioned and permissionless DLT systems. Example implementations are available for Hyperledger Fabric[9] and Corda [10] with further work underway on supporting other DLT platforms.

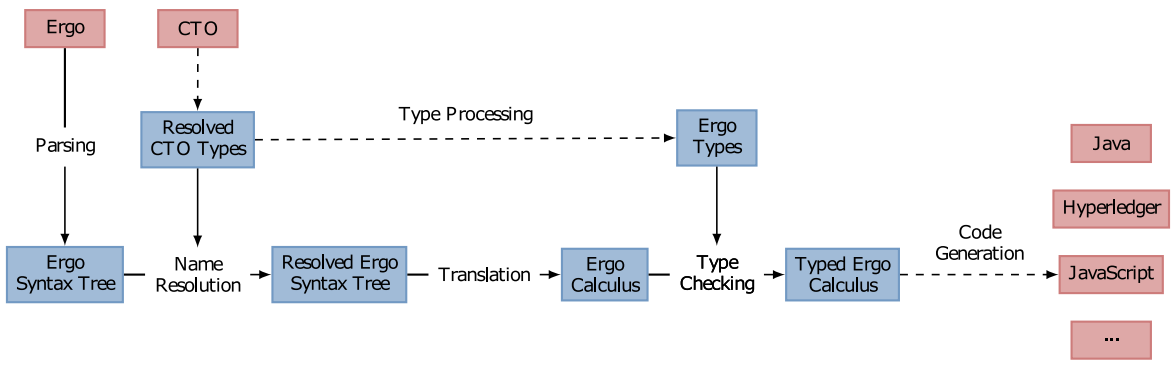

Figure 4: Ergo Compiler architecture [14]

## 2 Ergo Language Overview

The Ergo compiler (outlined in Figure 4) is written using the Coq[11] proof assistant, with parsing and support code written in OCaml[12]. It makes extensive use of the Q*cert Auerbach et al. (2017) compiler for code generation and type checking. The Ergo compiler is distributed as an npm package for Node.js [13].

Design Principles Ergo is based on the following principles:

1. Ergo contracts have a class-like structure with clauses akin to methods
2. Ergo can handle types (concepts, transactions, assets etc) specified in Concerto models
3. Ergo borrows from strongly-typed functional programming languages: clauses have a well-defined type signature (input and output), they are functions without side effects
4. The compiler guarantees error-free execution for well-typed Ergo programs
5. Clauses and functions are written in an expression language with limited expressiveness (it allows conditional and bounded iteration)
6. Most of the compiler is written in Coq as a stepping stone for formal specification and verification

A partial extract from an Ergo implementation of the logic for an acceptance of delivery clause[15]

[6] https://docs.accordproject.org/docs/tutorial-nodejs.html
[7] https://aws.amazon.com/qldb/
[8] https://github.com/accordproject/aws-qldb-lambda
[9] https://docs.accordproject.org/docs/tutorial-hyperledger.html
[10] https://accordproject.org/news/smart-legal-contracts-on-corda/
[11] https://coq.inria.fr
[12] https://ocaml.org/
[13] https://nodejs.org/
[15] https://studio.accordproject.org

would be:

```
let status =
  if isAfter(now(),
   addDuration(received,
    Duration{ amount: contract.
    businessDays,
      unit: TemporalUnit.days}))
  then OUTSIDE_INSPECTION_PERIOD
  else if request.inspectionPassed
  then PASSED_TESTING
  else FAILED_TESTING
;
```

Ergo can be used in standalone script or embedded in the text of the legal agreement clauses using Ergo Expressions as in the example below: [16]

```
Each party hereby irrevocably agrees
that process may be served on it
in any manner authorized by the Laws of
{{%
 if address.country = US
   and getYear(now()) > 1959
 then "the State of " ++ address.state
 else "the Country of " ++ address.
    country
%}}

The average price of the products
    included
in this purchase order is
{{%
avg(foreach p in products return p.price
    )
%}}
```

Ergo has built in support for concepts relating to date comparisons and for representing time intervals as these are a recurring type of smart legal contract clause logic. As an example:

```
This lease was signed on {{signatureDate
    }},
and is valid for a {{leaseTerm}} period.
This lease will expire on
{{% addPeriod(signatureDate, leaseTerm)
    %}}`
```

As legal contracts represent parties to an agreement Ergo also has a foundation concept of a *Party* and *Organization*. The concept of a contract state and state changes over a contract lifecycle are also a core part of the language, with the state changing as a result of invocation of clauses or as a result of other function calls.

In addition to the ability to create custom concepts, Ergo introduces the concept of an Legal Obligation that may arise as part of a legal agreement. The *PaymentObligation* is a first order language type as many clauses may involve the need to represent payments due to regular agreed intervals or penalties for late payments as an example.

```
emit PaymentObligation{
  promisor: some(contract.BUYER),
  promisee: some(contract.SELLER),
  amount: MonetaryAmount{
    doubleValue : request.amount,
    currencyCode : USD
  }
};
```

Clauses are implemented as functions that are called in a familiar Request and Response invocation style passing input parameters and receiving response values with possible events such as obligations being emitted. The outputs from the clause invocation could be used to initiate events that are sent to a DLT solution, or alternatively, to a banking payment gateway in the case of a *PaymentObligation*.

## 3 Conclusion

We hope that this work will contribute towards deepening the ongoing collaboration between the legal industry and the Accord Project community, and encourage further development of templates, model and standardised Ergo code libraries.

In future work we intend to expand the breath and depth templates and models available and to support additional DLT based systems, including support for WebAssembly. Work is also progressing in improving the tool set for contract creation. As an example the Accord Project has been working on using an NLP based approach to automatically create Cicero datatypes automatically from the contract text. [17]

[16] https://docs.accordproject.org/docs/markup-ergo.html

[17] https://accordproject.org/news/gsoc-2021-automatic-identification-and-classification